\documentclass[prd,twocolumn,eqsecnum,showpacs,amsfonts,amssymb]{revtex4}

\usepackage{graphicx}

\usepackage{bm}

\newcommand{\beq}{\begin{equation}}
\newcommand{\eeq}{\end{equation}}
\newcommand{\beqs}{\begin{eqnarray}}
\newcommand{\eeqs}{\end{eqnarray}}

\begin{document}

\title{Zeros of the Potts Model Partition Function on Sierpinski Graphs}  

\author{Shu-Chiuan Chang$^a$ and Robert Shrock$^b$}

\affiliation{(a) \ Physics Department \\ 
National Chen Kung University,  \\
Tainan 70101, Taiwan}

\affiliation{(b) \ C. N. Yang Institute for Theoretical Physics \\
Stony Brook University \\
Stony Brook, NY 11794 }

\begin{abstract}

We calculate zeros of the $q$-state Potts model partition function on
$m$'th-iterate Sierpinski graphs, $S_m$, in the variable $q$ and in a
temperature-like variable, $y$. We infer some asymptotic properties of the loci
of zeros in the limit $m \to \infty$ and relate these to thermodynamic
properties of the $q$-state Potts ferromagnet and antiferromagnet on the
Sierpinski gasket fractal, $S_\infty$.

\end{abstract}


\maketitle

\section{Introduction} 

Studies of iterated transformations and fractals have produced many interesting
results in physics and mathematics \cite{fractalrev}.  There have been several
analyses of spin models and critical phenomena on fractals; some early works
are \cite{bo}-\cite{ddi}.  Analyses of zeros of partition functions of spin
models in temperature-dependent Boltzmann variables (Fisher zeros) and
field-dependent variables (Yang-Lee zeros) have been performed, often for the
Ising model, and, in some cases, also the Potts model
\cite{ddi}-\cite{blr}.  A different but related area of interest is
vertex colorings of one-parameter families of graphs $G_m$ that lead to
fractals in the limit $m \to \infty$, denoted $G_\infty \equiv \lim_{m \to
\infty} G_m$.  For an arbitrary graph $G$, the number of ways of assigning $q$
colors to the vertices of $G$ subject to the condition that no two adjacent
vertices have the same color is enumerated by the chromatic polynomial,
$P(G,q)$. 

Here we report calculations of chromatic polynomials and their zeros for $m$'th
iterates of Sierpinski graphs, denoted $S_m$, and discuss their asymptotic
behavior in the limit $m \to \infty$, where $S_\infty$ is the Sierpinski gasket
fractal. More generally, we also analyze the zeros of the $q$-state Potts model
partition function in the complex $q$ plane at finite temperature for both the
antiferromagnetic (AFM) and ferromagnetic (FM) cases.  We also give some
results for zeros in the plane of a temperature-like Boltzmann variable (i.e.,
Fisher zeros).

We first recall some relevant background.  A graph $G=(V,E)$ is defined by its
set $V$ of vertices (sites) and its set $E$ of edges (bonds).  We denote
$n=n(G)=|V|$ and $e(G)=|E|$ as the number of vertices and edges of $G$. At
temperature $T$, the (zero-field) Potts model partition function is given by $Z
= \sum_{ \{ \sigma_n \} } e^{-\beta {\cal H}}$, with the Hamiltonian
\cite{wurev} 
\beq
{\cal H} = -J \sum_{\langle i j \rangle} \delta_{\sigma_i, \sigma_j} \ , 
\label{ham}
\eeq
where $i$ and $j$ label vertices of $G$, $\sigma_i$ are
classical spin variables on these vertices, taking values in the set
$I_q = \{1,...,q\}$, $\beta = (k_BT)^{-1}$, and $\langle
i j \rangle$ label pairs of adjacent vertices, or equivalently, edges in $E$.
It will be convenient to introduce the notation
\beq
K = \beta J \ , \quad h = \beta H \ , \quad y = e^K, \quad v=y-1 \ . 
\label{kdef}
\eeq
Thus, the physical ranges of $v$ are $v \ge 0$ for the Potts ferromagnet 
and $-1 \le v \le 0$ for the Potts antiferromagnet. 

A spanning subgraph of $G$ is $G' = (V,E')$ with $E' \subseteq E$. The number 
of connected components of $G'$ is denoted $k(G')$.  The Potts
partition function can be expressed as the sum over spanning subgraphs
\cite{fk} 
\beq
Z(G,q,v) = \sum_{G' \subseteq G} v^{e(G')} \ q^{k(G')} \ .
\label{cluster}
\eeq
As is evident from this expression, the partition function of the $q$-state
Potts model is a polynomial in $q$ and $v$ with positive integer coefficients.
For the Potts FM, this expression allows one to generalize $q$ from ${\mathbb
Z}_+$ to ${\mathbb R}_+$ while maintaining a positive Gibbs measure.
Eq. (\ref{cluster}) also allows one to generalize $q$ and $v$ to complex
numbers, as is necessary to analyze zeros of $Z(G,q,v)$.  For real $v$, the
zeros of $Z(G,q,v)$ in the $q$ plane are invariant under complex conjugation,
and for real $q$, the zeros of $Z(G,q,v)$ in the $v$ or $y$ plane are invariant
under complex conjugation.

For the Potts AFM, $T \to 0$ means $K \to -\infty$ and thus $v \to -1$. As is
clear from Eq. (\ref{ham}), the only spin configurations that contribute to
$Z(G,q,v)$ in the limit $v \to -1$ are those for which the spins on adjacent
vertices are different.  This proves that $Z(G,q,-1)=P(G,q)$, where $P(G,q)$ is
the chromatic polynomial, which, for $q \in {\mathbb Z}_+$, counts the number
of ways of assigning $q$ colors to the vertices of $G$ subject to the condition
that no two adjacent vertices have the same color (called proper $q$-colorings
of $G$). The minimum integer $q$ that allows a proper $q$-coloring of $G$ is
the chromatic number, $\chi(G)$.  

Besides its intrinsic interest in mathematical graph theory, the chromatic
polynomial is important for physics because of its connection with ground-state
entropy.  For a family $G_m$, in the limit $m \to \infty$ (and hence $n \to
\infty$), we define 
\beq
W(G_\infty,q) = \lim_{n \to \infty} P(G_m,q)^{1/n}  \ . 
\label{w}
\eeq
For a set of special $q$ values, $\{ q_s \}$, the limits $n \to \infty$ and $q
\to q_s$ do not, in general, commute for $P(G,q)^{1/n}$, so one must specify
the order of limits in defining $W(G_\infty,q)$ \cite{w,a}.  The set $\{ q_s
\}$ includes $q=0, \ 1, \ 2$. For our purposes here, we define the order as $q
\to q_s$ first and then $n \to \infty$.  For a large class of $G_m$, if $q$ is
sufficiently large, then the number of proper $q$-colorings of $G_m$ grows
exponentially with $m$ and $n$, so that the Potts AFM has nonzero ground-state
entropy per vertex on $G_\infty$, $S(G_\infty,q) = k_B \ln[W(G_\infty,q)] \ne
0$.

The Sierpinski graph $S_m$ is a planar graph constructed in the following
iterative (i.e., hierachical) manner \cite{fractalrev}. One starts at stage
$m=0$ with a triangle, then inserts vertices on the three edges and joins these
with new edges to get $S_1$.  The central triangle is left unchanged, and at
the next stage, $m=2$, one repeats this procedure with the three other interior
triangles, and so forth for higher $m$. The resultant graph at stage $m$ is
denoted $S_m$.  We note some basic properties of $S_m$. 
The numbers of vertices and edges on $S_m$ are, respectively,
\beq
n(S_m) = \frac{3(3^m+1)}{2} \ , \quad e(S_m)=3^{m+1} \ .
\label{nesm}
\eeq
The degree of a vertex is defined as the number of edges that connect to it.
The graph $S_m$ has three vertices of degree two, namely the ones at the three
corners of the outermost triangle, while the other $(3/2)(3^m-1)$ vertices have
degree 4. For a graph $G$ with vertices of different degrees, it is convenient
to define an effective or average vertex degree. This is
\beq
\kappa_{eff}(G) = \frac{2e(G)}{n(G)} \ . 
\label{kappaeff}
\eeq
Defining $\kappa_{eff}(G_\infty) \equiv \lim_{n \to \infty} 2e(G)/n(G)$, we
have
\beq
\kappa_{eff}(S_\infty) = 4 \ . 
\label{sgkappaeff}
\eeq
The chromatic number is easily shown to be $\chi(S_m)=3$, and $S_m$ is
tripartite, i.e.
\beq
P(S_m,3)=3! 
\label{psmq3}
\eeq
The girth of a graph $G$, $g(G)$, is defined
as the number of edges in a minimal-length closed (non-backtracking) path on
$G$.  This is $g(S_m)=3$ for all $m$.  The fractal dimension $d_f$ of
$S_\infty$ is \cite{fractalrev} $d_f(S_\infty)= \ln 3/\ln 2 \simeq 1.585$. 
Some graphical properties of $S_m$, including the numbers of spanning trees,
spanning forests, dimer coverings, and Hamiltonian walks, have been studied
recently in \cite{stsf}. 


\section{Chromatic Polynomial and Zeros} 

As with other hierarchical families of graphs, one can calculate
$Z(S_{m+1},q,v)$ from $Z(S_m,q,v)$ \cite{borjan93,andrade93}.  We have done
this up to $m=6$. We first discuss the special case of $P(S_m,q)=Z(S_m,q,-1)$.
In general, if a graph $G$ contains a complete graph $K_r$ \cite{kr} as a
subgraph, then $P(G,q)$ contains the factor
$P(K_r,q)=\prod_{j=0}^{r-1}(q-j)$. Since $S_m$ contains one or more triangles
$K_3$ as subgraph(s), $P(S_m,q)$ contains the factor $q(q-1)(q-2)$.  From
Eq. (\ref{psmq3}), it follows that
\beq
W(S_\infty,3)=1 \ . 
\label{wsgq3}
\eeq
For $q=4$, from our calculations up to $m=5$, we infer that 
\beq
P(S_m,4) = 2^{3(3^{m-1}+1)} \cdot 3^{3^{m-1}} \ . 
\label{psmq4}
\eeq
In the limit $m \to \infty$, this gives $W(S_\infty,4) = 2^{2/3} \cdot 3^{2/9}
= 2.026346..$, and hence the $q=4$ Potts antiferromagnet has a nonzero ground
state entropy \cite{andrade93} $S(S_\infty,4)=(2/3)k_B[\ln 2 + (1/3)\ln 3] =
(0.706234..)k_B$.

\begin{figure}
\begin{center}
\includegraphics[height=6cm]{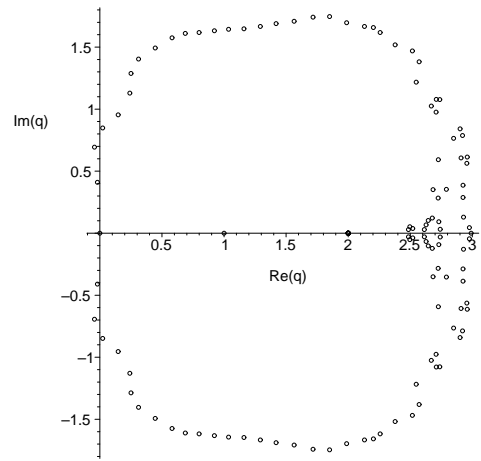}
\end{center}
\caption{\footnotesize{Zeros of the chromatic polynomial $P(S_4,q)$ 
in the $q$ plane (123 zeros).}}
\label{psgm4zeros}
\end{figure}

\begin{figure}
\begin{center}
\includegraphics[height=6cm]{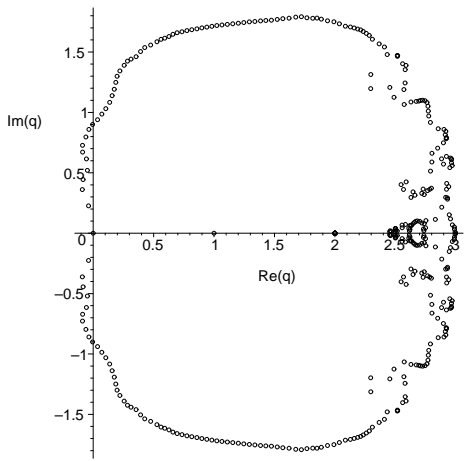}
\end{center}
\caption{\footnotesize{Zeros of the chromatic polynomial $P(S_5,q)$ 
in the $q$ plane (366 zeros).}}
\label{psgm5zeros}
\end{figure}

The calculations of $P(S_m,q)$ for $m=0$ and $m=1$ are elementary and yield
$P(S_0,q) =q(q-1)(q-2)$ and $P(S_1,q) = q(q-1)(q-2)^4$. We have computed the
zeros of $P(S_m,q)$ in the $q$ plane for $m$ up to 5.  We show plots of these
chromatic zeros for $m=4$ and $m=5$ in Figs.  \ref{psgm4zeros} and
\ref{psgm5zeros}.  For $m=0$ and $m=1$, the only chromatic zeros are real and
occur at $q=0, \ 1, \ 2$. For $m=1$, the chromatic zero at $q=2$ has
multiplicity 4 and for the higher $m$ values, we find that this zero has
multiplicity $3m$. For all $m$ values that we have studied, we find that the
only real chromatic zeros of $S_m$ are at $q=0$ and $q=1$ (both with
multiplicity 1), $q=2$ with the multiplicities given above, and, for $m \ge 2$
at a value slightly less than $q=3$, which monotonically increases toward $q=3$
as $m$ increases.  Specifically, for $m=2,3,4,5$, this largest real zero of
$P(S_m,q)$ occurs at 2.865331, 2.956063, 2.990362, 2.9977175, respectively.


As in earlier work \cite{w}, we denote the continuous accumulation set of
chromatic zeros as $m \to \infty$ as ${\cal B}_q$ and the maximal point at
which ${\cal B}_q$ intersects the real axis as $q_c$. As is evident in
Figs. \ref{psgm4zeros} and \ref{psgm5zeros}, the outermost zeros are consistent
with the inference that in the $m \to \infty$ limit a subset of zeros forms a
closed loop on ${\cal B}_q$ that crosses the real axis at $q=0$ and at a
maximal point
\beq
q_c(S_\infty)=3 \ . 
\label{qcsg}
\eeq
For a one-parameter family of graphs $G_m$, the locus ${\cal B}_q$ is a slice,
for constant $v=-1$, of the locus in the ${\mathbb C}^2$ space formed by the
merging of zeros of $Z(G_m,q,v)$ \cite{a}.  In general, if $q_c(G_\infty) \in
{\mathbb Z}_+$, this corresponds to the property that for this value of $q$,
${\cal B}_v$ passes through $v=-1$, i.e., ${\cal B}_y$ passes through $y=0$,
signifying that the Potts AFM with $q=q_c(G_\infty)$ has a $T=0$ critical point
on $G_\infty$.  Eq. (\ref{qcsg}) is equivalent to the property that the 
$q=3$ Potts AFM is critical on $S_\infty$ at $T=0$. 

It is intriguing that, despite the fact that $S_\infty$ shares three properties
with the triangular lattice \cite{thermodynamiclimit}, namely (i) the girth
equal to 3, (ii) the tripartite property, and (iii) the property of containing
infinitely many triangle subgraphs, the value $q_c(S_\infty)=3$ is different
from the value $q_c(tri)=4$ for the triangular lattice.  Indeed, the value
$q_c(S_\infty)=3$ is the same as the value $q_c(sq)=3$ for the square lattice
\cite{wurev}.  We showed via exact calculations that $q_c=3$ for
infinite-length, finite-width self-dual strips of the square lattice
\cite{sdg}. The result $q_c(S_\infty)=3=q_c(sq)$ is also interesting in view of
the differences between $S_\infty$ and the square lattice.  Even before taking
the $m \to \infty$ limit, these differences are evident. Thus, as noted above,
$S_m$ has $\chi(S_m)=3$ so that $P(S_m,2)=0$ and is tripartite, whereas a
finite $L_x \times L_y$ section of the square lattice with free boundary
conditions or toroidal boundary conditions with even $L_x$ and $L_y$ is
bipartite, so for both of these families of graphs, denoted $sq[L_x \times
L_y,BC]$, $\chi=2$ and $P(sq[L_x \times L_y,BC],2)=2$.  Furthermore, $P(S_m,3)$
is a constant, namely $3!$, while $P(G[sq,L_x \times L_y,BC],3)$ grows
exponentially rapidly with $n=L_xL_y$.  One property that $S_\infty$ and the
square lattice have in common is that $\kappa_{eff}(S_\infty)=\kappa(sq)=4$.

We remark further on the pattern of chromatic zeros in Figs.  \ref{psgm4zeros}
and \ref{psgm5zeros}.  The zeros on the upper and lower part of the outer loop
exhibit a rather smooth appearance, although undulations on a sufficiently
small scale would not be visible with our numbers of zeros.  On the left side
there is a decreased density of zeros, and on the right side the pattern of
zeros has a more wavy and irregular structure. It is possible that some of the
irregularly distributed zeros on the right might form further continuous
substructures on ${\cal B}_q$ in the $m \to \infty$ limit.  One also sees zeros
in the interior that are consistent with the inference that in the $m \to
\infty$ limit they merge to form an inner closed loop on ${\cal B}_q$ that
crosses the real axis on the left at $q \simeq 2.62$ and on the right at $q
\simeq 2.74$.  We note that the left crossing point is equal, to within the
uncertainty, with $(3+\sqrt{5})/2 = 4\cos^2(\pi/5) = 2.6180...$. There is also
a substantial set of complex zeros scattered in an area whose center, on the
real axis, occurs at $q \simeq 2.5$; again, these could form further continuous
substructures as $m \to \infty$.


\section{General Zeros of $Z(S_m,q,v)$ in the $q$ Plane}

As noted above, the chromatic polynomial is equal to the partition function of
the Potts antiferromagnet at $T=0$, i.e., $v=-1$.  It is of interest to study
the zeros of the Potts partition function $Z(S_m,q,v)$ in the $q$ plane for
nonzero temperature in the antiferromagnetic interval $v \in (-1,0]$ and also
for the ferromagnetic interval $v > 0$. We report these results here.  For our
purposes, it will suffice to use our calculation of $Z(S_4,q,v)$.

\begin{figure}
\begin{center}
\includegraphics[height=6cm]{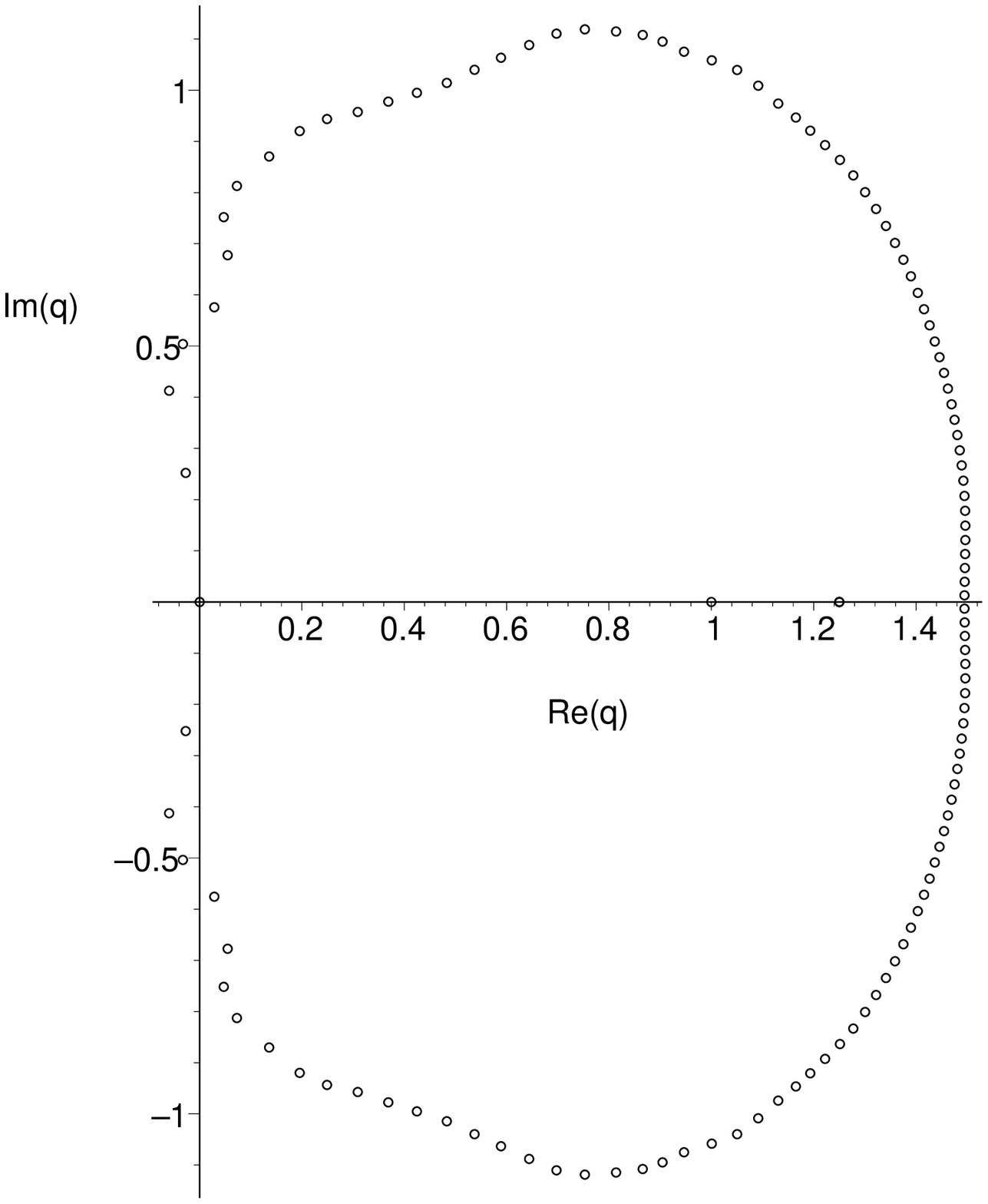}
\end{center}
\caption{\footnotesize{Zeros of $Z(S_4,q,v)$ in the $q$ plane for
$v=-0.5$.}}
\label{sgqplotvm0p5}
\end{figure}

In Fig. \ref{sgqplotvm0p5} we show zeros of $Z(S_4,q,v)$ in the $q$ plane for
the finite-temperature AFM value $v=-0.5$.  These zeros are consistent with the
inference that in the $m \to \infty$ limit, ${\cal B}_q$ forms a closed locus
that separates the $q$ plane into two regions, crossing the real axis at $q=0$
and a $q_c$ value that has decreased from its value of 3 at $v=-1$ to $\sim
1.5$.  (Strictly speaking, one cannot rule out the possibility of additional
very small sliver regions, as were found in \cite{wcy}, but we see no evidence
for them here.)  As was the case for regular infinite-length, finite-width
lattice strips (e.g., \cite{a,ta}), we find that $q_c$ decreases monotonically
from its value of 3 at $v=-1$ to 0 as $v$ increases from $-1$ to 0.  This is
understandable, since for the Potts model on an arbitrary graph $G$,
$Z(G,q,0)=q^n$, so that in the limit $v \to 0$, all of the zeros in $q$ occur
at the single value $q=0$ and the locus ${\cal B}_q$ degenerates to this point.
A notable structural change in these $q$-plane zeros that occurs as one
increases $v$ from $-1$ to $-0.5$ is that the zeros on the right side exhibit a
smoother appearance, without visible outer or inner protruberances.

\begin{figure}
\begin{center}
\includegraphics[height=6cm]{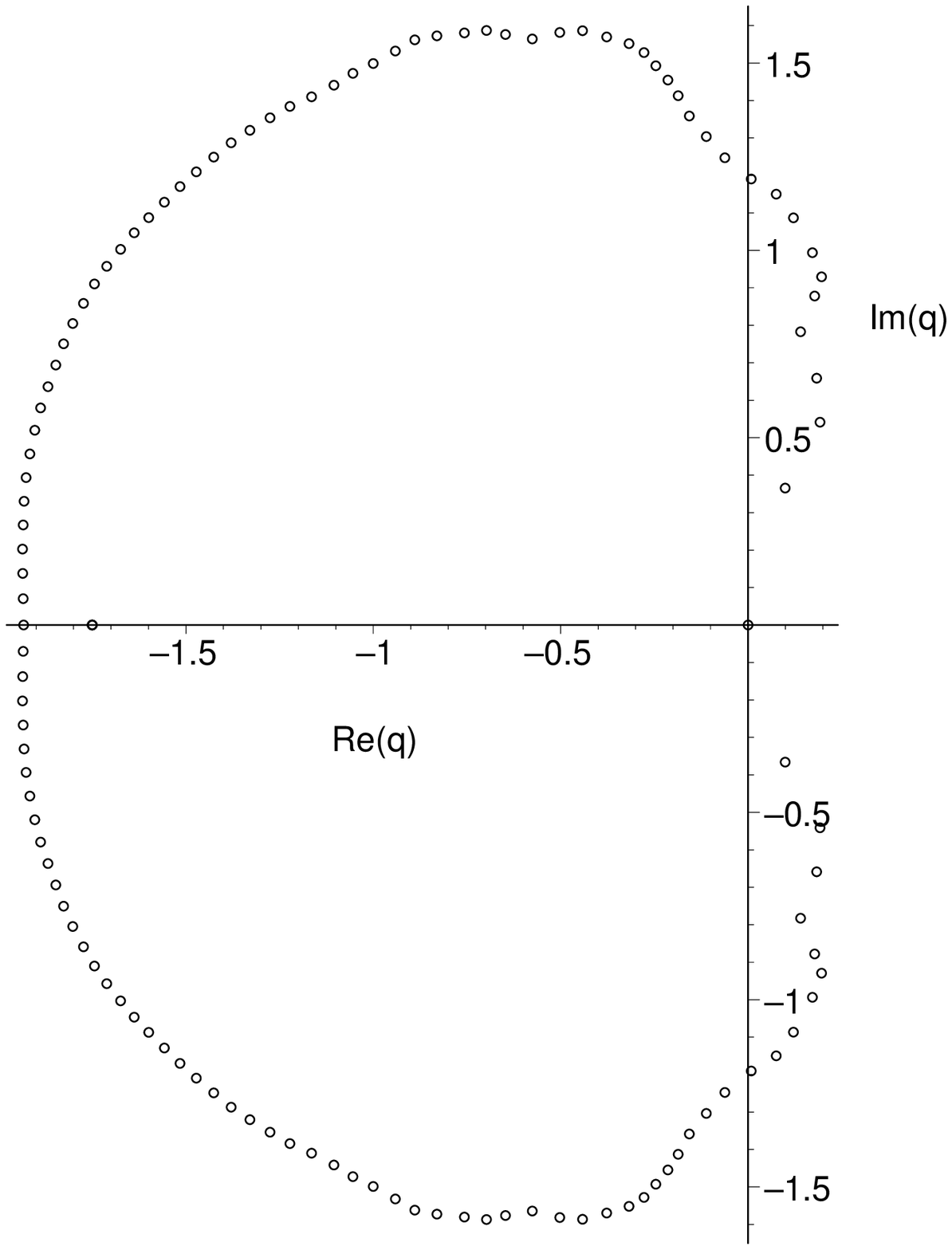}
\end{center}
\caption{\footnotesize{Zeros of $Z(S_4,q,v)$ in the $q$ plane for
$v=0.5$.}}
\label{sgqplotv0p5}
\end{figure}

We next consider the FM interval and show zeros of $Z(S_4,q,v)$ in $q$ for the
illustrative value $v=0.5$ in Fig. \ref{sgqplotv0p5}.  These are consistent
with the inference that for $m \to \infty$, ${\cal B}_q$ crosses the real axis
at a value $q \simeq -1.9$ and does not cross this axis at any positive
value. The latter result is in accord with the property that 
the $q$-state Potts FM on $S_\infty$ has no 
finite-temperature critical point \cite{gk,gam}.  Just as the density of zeros
on the left side decreases in the vicinity of $q=0$ in the plots for
$v=-1$ and $v=-0.5$, so also for this FM value, $v=0.5$, the density of zeros
decreases in the vicinity of $q=0$.  

%


\section{Zeros of $Z(S_m,q,v)$ in the $v$ Plane}

\begin{figure}
\begin{center}
\includegraphics[height=6cm]{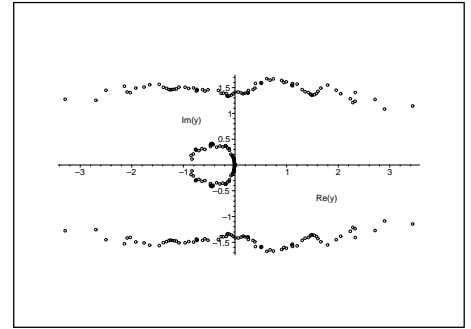}
\end{center}
\caption{\footnotesize{Zeros of $Z(S_4,q,v)$ in the $y$ plane for
$q=3$.}}
\label{sgyplotq3}
\end{figure}

\begin{figure}
\begin{center}
\includegraphics[height=6cm]{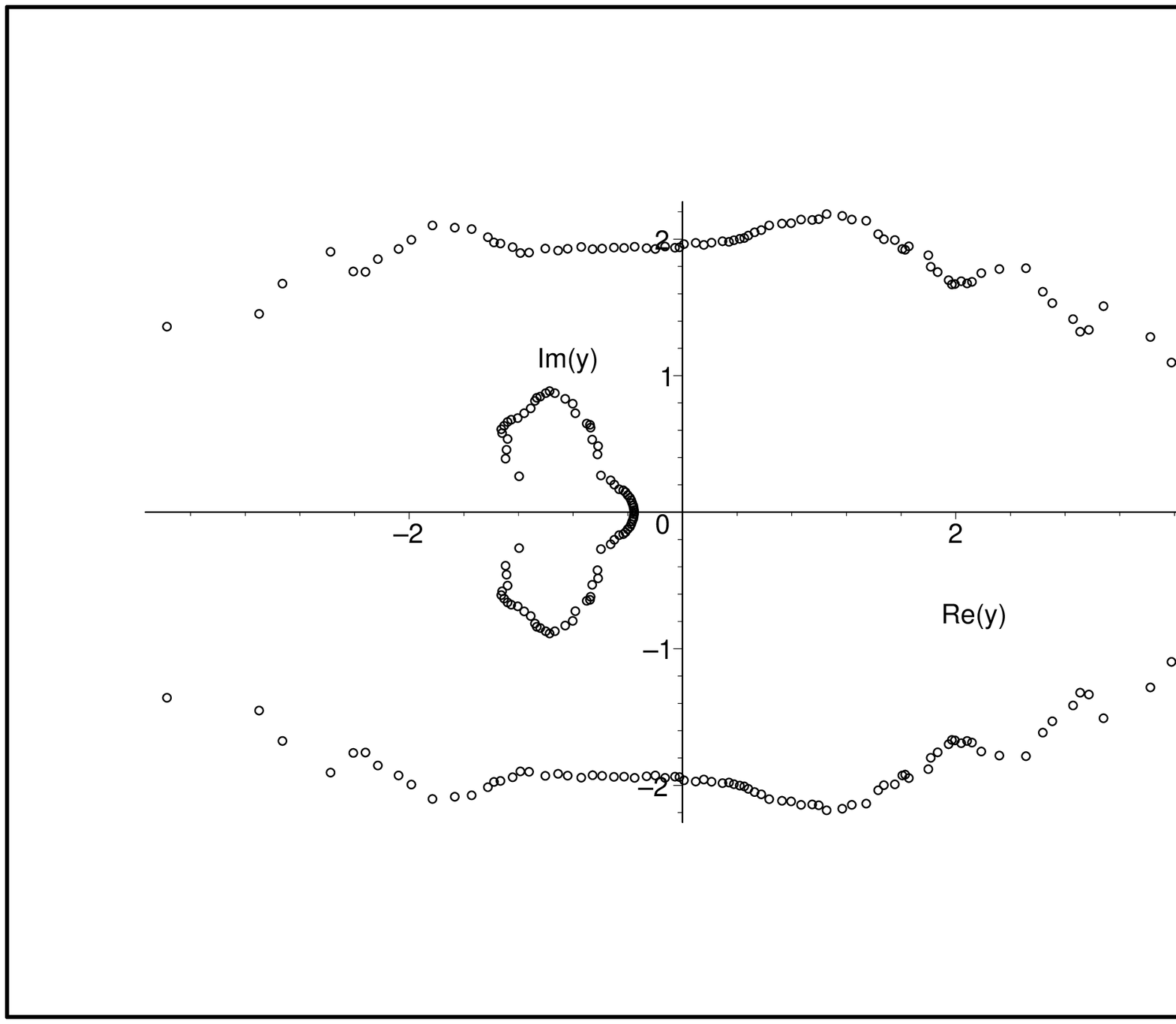}
\end{center}
\caption{\footnotesize{Zeros of $Z(S_4,q,v)$ in the $y$ plane for
$q=5$.}}
\label{sgyplotq5}
\end{figure}

\begin{figure}
\begin{center}
\includegraphics[height=6cm]{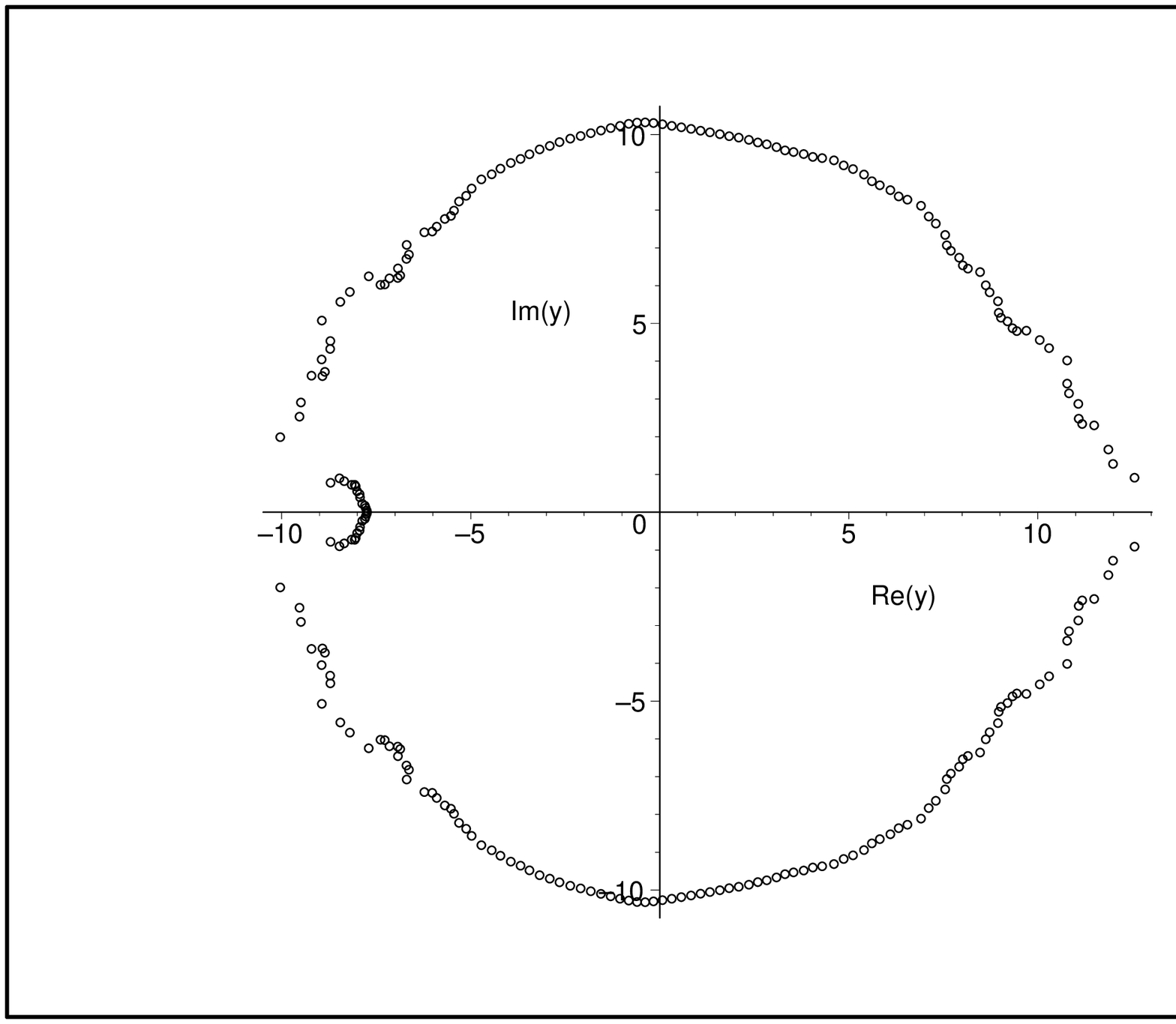}
\end{center}
\caption{\footnotesize{Zeros of $Z(S_4,q,v)$ in the $y$ plane for
$q=10^2$.}}
\label{sgyplotq100}
\end{figure}

\begin{figure}
\begin{center}
\includegraphics[height=6cm]{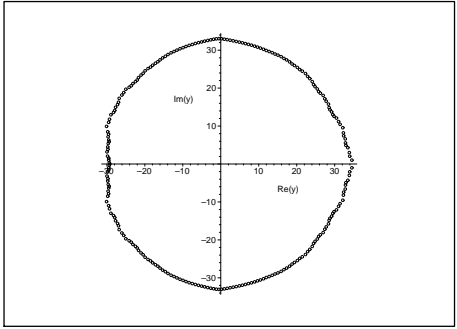}
\end{center}
\caption{\footnotesize{Zeros of $Z(S_4,q,v)$ in the $y$ plane for
$q=10^3$.}}
\label{sgyplotq1000}
\end{figure}

We have also calculated zeros of $Z(S_m,q,v)$ in the complex plane of a
temperature-like Boltzmann variable $v$, or equivalently, $y=v+1$ as a function
of $q$.  Plots of these zeros for $q=2, \ 3, \ 4$, were presented in
\cite{desimoi2009}, and we confirm these results.  Plots for larger values of
$q$ are also of interest, and we present these here.  We denote the continuous
locus formed by the accumulation set of these zeros in the limit $m \to \infty$
as ${\cal B}_y$. We show zeros of $Z(S_4,q,v)$ in the $y$ plane for $q=3, \ 5, 
10^2$, and $10^3$ in Figs. \ref{sgyplotq3}-\ref{sgyplotq1000}.


From our calculations of zeros of $Z(S_m,q,v)$ in $v$ for various $q$ and $m$,
we can draw a number of inferences concerning ${\cal B}_y$ for $S_\infty$.
First, for the Ising value $q=2$, ${\cal B}_y$ passes through the point
$1/y=0$, in agreement with the fact that the Ising FM has a $T=0$ critical
point on the Sierpinski gasket $S_\infty$, as evidenced from the divergence in
the correlation length as $T \to 0$ \cite{andrade93}.  In the $1/y$ plane,
${\cal B}_{1/y}$ forms roughly a figure-8 shape, crossing the imaginary $1/y$
axis at $\pm 0.7i$.  Second, equivalent to our conclusion from the chromatic
zeros that $q_c(S_\infty)=3$, we infer that for the $q=3$ Potts AFM on
$S_\infty$, ${\cal B}_y$ passes through $y=0$. Our results for $q_c(S_\infty)$
also imply that ${\cal B}_y$ does not pass through $y=0$ for the $q$-state
Potts antiferromagnet with $q > 3$. Our calculations of zeros in $y$ for $q >
3$ agree with this.  As $q$ increases beyond $q=3$, the inner loop of zeros
moves to the left.  As is evident in Fig.  \ref{sgyplotq1000}, for sufficiently
large $q$, there is no longer any visible inner loop.  As $q$ gets large, the
outer loop of zeros becomes more circular, and we infer that in the limit $m
\to \infty$ this is also true for ${\cal B}_y$. Since the $q$-state Potts
ferromagnet is critical at $T=0$ on $S_\infty$ \cite{gam}, it follows that
${\cal B}_y$ always passes through the point $1/y=0$.  As is evident from
Fig. \ref{sgyplotq100}, this means that as one approaches the real axis on
${\cal B}_y$, the locus deviates from the roughly circular pattern to pass
through $1/y=0$.  These inferences are in accord with our earlier work in
\cite{lq} (see also \cite{wulq}, where it was shown that the radius of this
approximately circular pattern of zeros is
\beq
|y| = q^{2/\kappa_{eff}} \quad {\rm as} \ \ q \to \infty \ . 
\label{yradius}
\eeq
Since $\kappa_{eff}$ approaches 4 on $S_m$ with increasing $m$, this radius
approaches $|y|=\sqrt{q}$.  Hence, this radius is approximately 10 for the case
$q=10^2$ in Fig. \ref{sgyplotq100} and $\sim 32$ for $q=10^3$ in
Fig. \ref{sgyplotq1000}, in agreement with our results shown in these figures. 

We have also carried out similar calculations for generalizations of the
Sierpinski gasket, including the generalization to three dimensions.  With
R. Roeder, we have performed an analogous study for another family of graphs,
the $m$'th iterates of the diamond hierarchical lattice, $D_m$, which lead, in
the $m \to \infty$ limit, to the fractal $D_\infty$.  We find that for $D_m$,
the chromatic zeros are considerably more distributed throughout regions of the
$q$-plane than is the case with $S_m$.  Our results will be reported elsewhere.

This research was partially supported by the grants NSC-100-2112-M-006-003-MY3
(S.-C. C.) and NSF-PHY-09-69739 (R.S.).


\begin{thebibliography}{99}

\bibitem{fractalrev}
%
Some reviews and monographs include B. B. Mandelbrot, {\it The Fractal Geometry
of Nature} (Freeman, San Francisco, 1983); D. Stauffer and A. Aharony, {\it
Introduction to Percolation Theory}, 2nd ed. (Taylor and Francis, London,
1991); A. Bunde and S. Havlin, eds. , {\it Fractal and Disordered Systems}
(Springer, Berlin, 1996); J. Milnor, {\it Dynamics in One Complex Variable},
2nd ed.  (Vieweg, Berlin, 2000); A. F. Beardon, {\it Iteration of Rational
Functions} (Springer, Berlin, 1991). 

\bibitem{bo}
A. N. Berker and S. Ostland, J. Phys. C {\bf 12} (1979) 4961. 

\bibitem{gam}
%
Y. Gefen, B. B. Mandelbrot, and A. Aharony, Phys. Rev. Lett. {\bf 45} (1979) 
855; Y. Gefen, A. Aharony, and B. B. Mandelbrot, J. Phys. A {\bf 16} (1983)
1267; J. Phys. A {\bf 17} (1984) 435; J. Phys. A {\bf 17} (1984) 1277. 

\bibitem{gk}
M. Kaufman and R. B. Griffiths, Phys. Rev. B {\bf 24} (1981) 496; 
R. B. Griffiths and M. Kaufman, Phys. Rev. B {\bf 26} (1982) 5022; 
M. Kaufman and R. B. Griffiths, Phys. Rev. B {\bf 30} (1984) 244. 

\bibitem{bzl}
P. Bleher and E. Zalys, Commun. Math. Phys. {\bf 120} (1989) 409; 
P. M. Bleher and M. Yu. Lyubich, Commun. Math. Phys. {\bf 141} (1991) 453;

\bibitem{bhanot}
%
G. Bhanot, H. Neuberger, and J. A. Shapiro, Phys. Rev. Lett. {\bf 53} (1984)
2277; B. Hu, Phys. Rev. Lett. {\bf 55} (1985) 2316; B. Lin and Z. R. Yang 
J. Phys. A {\bf 19} (1987) L49; Y.-K. Wu and B. Hu, Phys. Rev. A {\bf 35}
(1987) 1404. 

\bibitem{ddi}
%
B. Derrida, L. De Seze, and C. Itzykson, J. Stat. Phys. {\bf 33} (1983) 559;
B. Derrida, C. Itzykson, and J. M. Luck, Commun. Math. Phys. {\bf 94} (1984) 
115. 

\bibitem{southern}
B. W. Southern and M. Kne\v{z}evi\'c, Phys. Rev. B {\bf 35} (1987) 5036.

\bibitem{borjan93}
%
Z. Borjan, M. Kne\v{z}evi\'c, and S. Milo\v{s}evi\'c, 
Phys. Rev. B {\bf 47} (1993) 144. 
Our calculations agree with this paper except that the first term on the
right-hand side of Eq. (4) should be $Z_4^3$ rather than $Z_4^2$. Also, the
term $q_3$ in Eq. (2) should be $q^3$. 

\bibitem{andrade93}
R. F. S. Andrade, Phys. Rev. B {\bf 48} (1993) 16095. 

\bibitem{qiao_etal}
%
Z. R. Yang, Phys. Rev. E {\bf 49} (1994) 2457; 
L. de Silva et al., Phys. Rev. B {\bf 53} (1996) 6345; 
J. Qiao and Y. Li, Commun. Math. Phys. {\bf 222} (2001) 319; 
T.-M. Liaw, M.-C. Huang, Y.-L. Chou, and S. C. Lin, Phys. Rev. E {\bf 65}
(2002) 066124; 
Y.-L. Chou and M.-C. Huang, Phys. Rev. E {\bf 67} (2003) 056109;
M.-C. Huang, Y.-P. Luo, and T.-M. Liaw, Phys. Lett. A {\bf 320} (2003) 180;
J. Gao and J. Qiao, Phys. Lett. A {\bf 355} (2006) 167. 

\bibitem{desimoi2009}
J. De Simoi, J. Phys. A {\bf 42} (2009) 095001, 095002. 

\bibitem{blr}
P. Bleher, M. Lyubich, and R. Roeder, arXiv:1009.4691; arXiv:1107.5764.

\bibitem{stsf}
%
S.-C. Chang and L.-C. Chen, J. Stat. Phys. {\bf 126} (2007) 649;
J. Stat. Phys. {\bf 131} (2008) 631; 
Discrete Math. Theor. Comput. Sci. {\bf 12:3} (2009) 151; 
J. Math. Phys. {\bf 52} (2011) 023301. 

\bibitem{wurev}
F. Y. Wu, Rev. Mod. Phys. {\bf 54} (1982) 235. 

\bibitem{fk}
C. M. Fortuin and P. W. Kasteleyn, Physica {\bf 57} (1972) 536. 

\bibitem{w}
R. Shrock and S.-H. Tsai, Phys. Rev. E {\bf 55} (1997) 5165. 

\bibitem{a}
R. Shrock, Physica A {\bf 283} (2000) 388. 

\bibitem{kr}
%
The complete graph $K_r$ is defined as the graph with $r$ vertices such that
each vertex is connected by (single) edges to each of the other vertices, so
$e(K_r)={r \choose 2}$. 

\bibitem{thermodynamiclimit}
%
Here and below, when we refer to a regular lattice $\Lambda$ (square, 
triangular, etc.), the thermodynamic limit of $\Lambda$ is understood. 

\bibitem{sdg}
S.-C. Chang and R. Shrock, Physica A {\bf 301} (2001) 301; 
Phys. Rev. E {\bf 64} (2001) 066116. 

\bibitem{wcy}
R. Shrock and S.-H. Tsai, Phys. Rev. E {\bf 60} (1999) 3512; 
Physica A {\bf 275} (2000) 429. 

\bibitem{ta}
S.-C. Chang and R. Shrock,  Physica A {\bf 286} (2000) 189. 

\bibitem{lq}
S.-C. Chang and R. Shrock, Int. J. Mod. Phys. B {\bf 21} (2007) 979. 

\bibitem{wulq}
H. Y. Huang and F. Y. Wu, Int. J. Mod. Phys. B {\bf 11} (1997) 121. 

\end{thebibliography}
\end{document}